\documentstyle[twocolumn,prd,aps,epsf]{revtex}
\begin{document}
\draft
\title{
Equivalence of renormalized covariant and light-front perturbation theory.\\
I$\!$I.$\;\;$ Transverse divergences in the Yukawa model}
\author{N. C. J. Schoonderwoerd and
        B. L. G. Bakker\\
  Department of Physics and Astronomy, Vrije Universiteit, Amsterdam,
  The Netherlands 1081~HV\\}
\date{January 29 1998, revised 19 March 1998}
\maketitle
\def \ftri{\raisebox{-.5cm}{\epsfxsize=.9cm \epsffile[0 0 63 90 ]{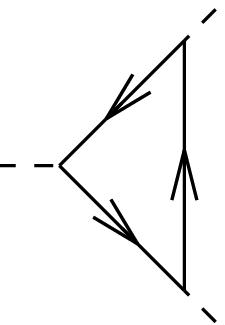}}}
\def \ftrix{\raisebox{-.8cm}{\epsfxsize=1.3cm \epsffile{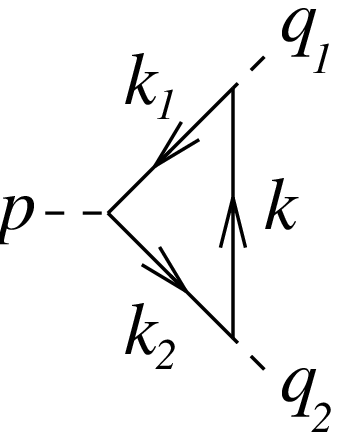}}}
\def \ftriapp{\raisebox{-.4cm}{\epsfxsize=1.2cm \epsffile{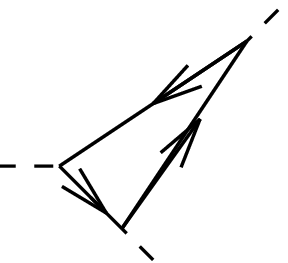}}}
\def \ftriaip{\raisebox{-.5cm}{\epsfxsize=1.2cm \epsffile{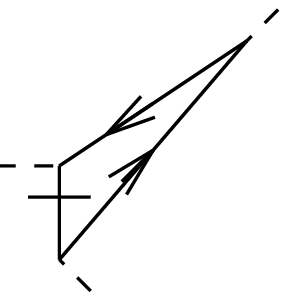}}}
\def \ftriapi{\raisebox{-.4cm}{\epsfxsize=1cm \epsffile{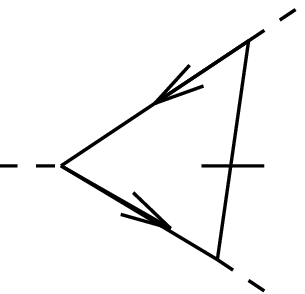}}}
\def \ftribpp{\raisebox{-.6cm}{\epsfxsize=1.2cm \epsffile{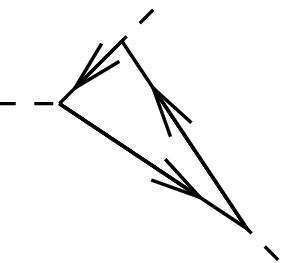}}}
\def \ftribip{\raisebox{-.6cm}{\epsfxsize=1.2cm \epsffile{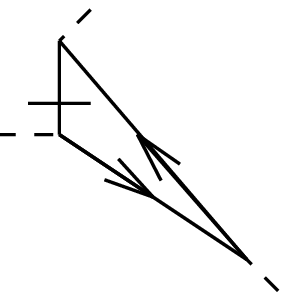}}}
\def \ftribpi{\raisebox{-.55cm}{\epsfxsize=1cm \epsffile{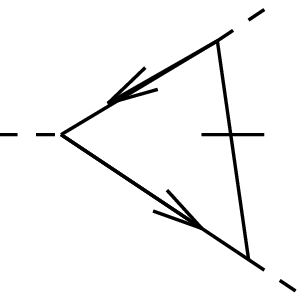}}}
\def \obe{\epsfxsize=.9cm \raisebox{-.6cm}{\epsffile{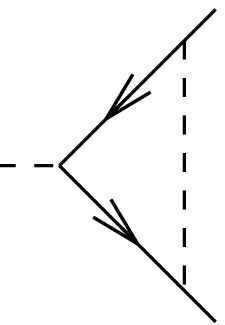}}}
\def \obex{\epsfxsize=1.3cm \raisebox{-.8cm}{\epsffile{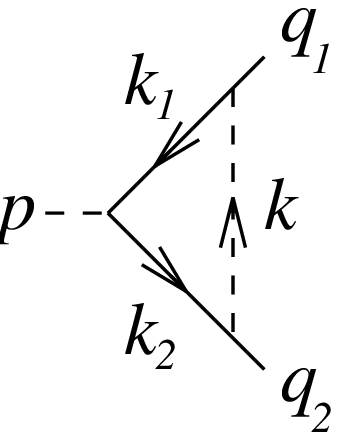}}}
\def \obeba{\epsfxsize=.9cm\raisebox{-.6cm}{\epsffile{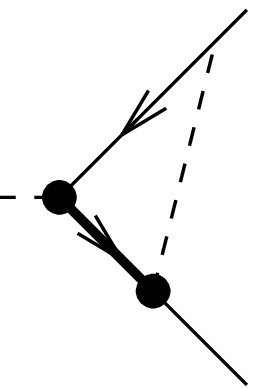}}}
\def \obebb{\epsfxsize=.9cm\raisebox{-.6cm}{\epsffile{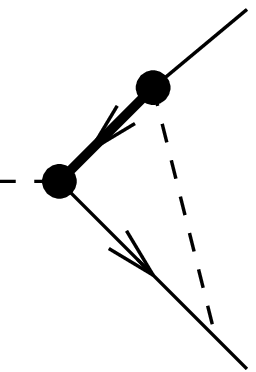}}}
\def \obepa{\epsfxsize=.9cm\raisebox{-.6cm}{\epsffile{obep1.eps}}}
\def \obepb{\epsfxsize=.9cm\raisebox{-.6cm}{\epsffile{obep2.eps}}}
\def \obeia{\epsfxsize=.9cm\raisebox{-.6cm}{\epsffile{obei1.eps}}}
\def \obeib{\epsfxsize=.9cm\raisebox{-.6cm}{\epsffile{obei2.eps}}}
\def \obeii{\epsfxsize=.9cm\raisebox{-.6cm}{\epsffile{obeii.eps}}}
\def \slash#1{\not\! #1}
\def \r#1{(\ref{#1})}

\begin{abstract}
Light-front dynamics can only become a viable alternative to the covariant
approach if doubts about its covariance can be taken away.  As a minimal
requirement we take that the physical quantities calculated with light-front
perturbation theory are the same as those obtained using covariant perturbation
theory. If this situation occurs, we use the word equivalent to characterize
it.  For quantities that involve the calculation of superficially convergent
diagrams, proofs of equivalence exist. For some types of divergent diagrams the
proof of equivalence is complicated. Here we deal with diagrams with transverse
divergences.  Our method is based on minus regularization, which is inspired on
BPHZ regularization. In a calculation using numerical methods we show how
to obtain a rotationally invariant amplitude for two triangle diagrams
contributing to the decay of a scalar boson in the Yukawa model. It concludes
our proof of equivalence of covariant and light-front perturbation theory.
Accepted for publication in Phys.Rev.D
\end{abstract}

\section{Introduction}
\label{secintro}
Covariant field theory is the formalism of choice to describe
situations where creation and annihilation of particles are important
and where typical velocities are comparable to the velocity of light.
If the interactions are sufficiently weak, perturbation theory is
usually applied and gives in many cases extremely accurate answers.
However, in the case of strong interactions, or when bound states are
considered, nonperturbative methods must be developed. Light-front
quantization \cite{Dir49} is a Hamiltonian method in which the
light-like variable $x^+ = (x^0 + x^3)/\sqrt{2}$ plays the role of
time, and is therefore referred to as light-front time. This method has
found many applications since it was conceived.  Still, some problems
of a fundamental nature remained.  One that we are particularly
interested in is the question of whether full covariance can be maintained
in the Hamiltonian formulation, which is of course not manifestly
covariant.  A partial answer can be obtained in perturbation theory.
Then the  problem can be reformulated as follows:  can one prove that
light-front perturbation theory produces the same values of the
S-matrix elements as covariant perturbation theory?  If the answer to
this question is affirmative, then we use the word equivalent to
describe the situation.

The present paper is concerned with one aspect of this problem, viz the
treatment of transverse divergences in a simple model: the Yukawa model
with spin-1/2 fermions, spin-0 bosons and a scalar coupling.

\subsection{$k^-$-integration and equivalence}

In the work we did before, we used the method of Kogut and Soper
\cite{KS70} to define light-front perturbation theory. This method
defines light-front time-ordered ($x^+$-ordered) amplitudes by
integration of the integrand of a covariant diagram, say

\begin{equation} \label{1500} F(q) = \int {\rm d}^4k\;\;
I(q;k), \end{equation}

\noindent
over the light-front energy variable $k^- = (k^0 - k^3)/\sqrt{2}$.  In
this paper, $q$ always denotes the external momenta and $k$ the loop
momentum. We can also write Eq.~\r{1500} using light-front coordinates:
\begin{equation} 
F(q)=\int \! {\rm d}k^+ {\rm d}^2k^\perp \int
{\rm d}k^- I(q^-\!\!,q^+\!\!,q^\perp;k^-\!\!,k^+\!\!,k^\perp).
\end{equation}
Next, one expresses the integral over $k^-$, using Cauchy's formula, as
a sum of residues.  One arrives in this way at an expression that can
be interpreted, possibly after recombination of the terms in this sum,
as the splitting of the covariant amplitude $F(q)$ into a sum
of noncovariant but light-front time-ordered amplitudes.

\begin{figure}
\epsfxsize=8.5cm \epsffile[ 0 0 531 189   ]{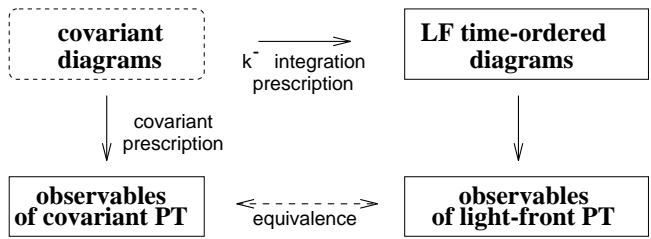}
\vspace{.2cm}
\caption{The ``ideal'' case:
\label{figeen}
Outline of our proof of equivalence of light-front (LF) and covariant
perturbation theory (PT) for convergent diagrams. The dashed
box indicates an ill-defined object.}
\end{figure}

This procedure, sometimes called naive light-cone quantization,
has been in principle known since the early work of Kogut and Soper~\cite{KS70}.
For convergent diagrams, it is nicely pictured in Fig.~\ref{figeen}.
 
The covariant diagram in Fig.~\ref{figeen} is an ill-defined object
and needs some prescription to give it a definite meaning.
For example, the measure of the Minkowskian integration
is not positive definite.  The covariant prescription involves
introduction of Feynman parameters to complete the squares in the
denominator, the removal of terms odd in the loop momentum $k$ and the
Wick rotation to obtain a Euclidian integral.

It has been the work of Ligterink and Bakker \cite{LB95b} that proves in detail
that the rules for constructing
light-front time-ordered diagrams, explained in many articles
\cite{KS70,LB80}, are correct upon using the $k^-$-integration
prescription.  They were the first authors to give a systematic
derivation of all the different time-ordered diagrams corresponding to
a given covariant amplitude, for any number of particles involved.  If
the $k^-$-integral is convergent and the corresponding covariant
diagram is also superficially convergent, then what remains can be
written in terms of well-defined, convergent Euclidian integrals.

When the $k^-$-integration is divergent, the prescription must be altered.
Naive light-front quantization fails in this case and one must first find a way
to regulate the $k^-$-integrals. We proposed in a previous paper \cite{SB98a}
a regularization that maintains covariance. There we showed that the
longitudinal divergences give rise to so-called forced instantaneous loops
(FILs) and we showed how to deal with them such that covariance is maintained.
This method was also applied
to the Yukawa model containing spin-$1/2$ and spin-$0$ particles.
We were able to regularize the $k^-$-integrals for the diagrams with one
loop. However, in order to show full equivalence to the
covariant calculation one needs to compute the full integral including
the integrations over $k^+$ and $k^\perp$.
 
\subsection{Ultraviolet and transverse divergences}
 
\begin{figure}
\epsfxsize=8.5cm \epsffile[ 0 0 526 193]{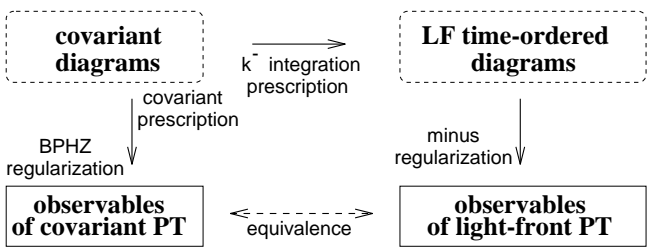}
\vspace{.2cm}
\caption{Outline of our proof of equivalence for diagrams with ultraviolet
divergences. \label{figtwee} Dashed boxes indicate ambiguously 
defined objects.}
\end{figure}

Even after the usual procedure has been followed, the covariant integral
can still be
ultraviolet divergent. Ligterink and Bakker did not only discuss diagrams that
are superficially convergent, but also what to do in cases
where the covariant diagram is divergent. Their method of regularizing
divergent diagrams, minus regularization \cite{LB95a}, is also used in the
present paper.  A scheme for the equivalence of ultraviolet
divergent diagrams is given in Fig.~\ref{figtwee}.  
 
Several techniques are available to remove the ultraviolet divergences,
not involving the $k^-$- integration. They remain
in the light-front time-ordered diagrams as divergences of the
integrals over the transverse momenta.  Therefore these diagrams are
also ill-defined, as indicated by the dashed box in Fig.~\ref{figtwee}.
A problem is that many of the techniques which are used to regularize
covariant diagrams have limited use for light-front time-ordered diagrams. For
example, one cannot use dimensional regularization for the
longitudinal divergences.  Still, it is common to apply it to the
transverse divergences.  The strength of the regularization scheme we
use, minus regularization, is that it does not discriminate between
transverse and longitudinal divergences.  Minus regularization
is based on the Bogoliubov-Parasiuk-Hepp-Zimmerman (BPHZ) method of
regularization \cite{Hep66,HZ66,Zim68,Zim69,CK82}.  In their paper,
Ligterink and Bakker applied minus regularization to three self-energy diagrams.
Our contribution is to extend their method to more complicated diagrams
and prove that there is a one-to-one relation between minus
and BPHZ regularization, such that the physical observables found using
light-front perturbation theory exactly match those found in covariant
perturbation theory.

In the Yukawa model there are five covariant diagrams with ultraviolet divergences.
The boson and the fermion self-energy were
discussed in our previous article on longitudinal divergences.
Minus regularization was applied and simultaneously removed the longitudinal
and the transverse divergences. Equivalence was established.

In two cases we were not able to either find an answer in the
literature or produce ourselves full analytic results for the
integrals involved; so we had to resort to numerical integration. 
In this paper we discuss
these two diagrams: the one-boson exchange correction to the
boson-fermion-fermion vertex and the fermion loop with three
external boson lines. 
The first one was considered by Burkardt and Langnau \cite{BL91}, who 
stated that naive light-cone quantization leads to a violation of 
rotational invariance of the corresponding S-matrix elements and
found that invariant results can be obtained using noncovariant counterterms.
Here we show that no violation of rotational invariance occurs
if our method of regularization is applied. Furthermore, our results
for the light-front time-ordered diagrams sum up to the covariant amplitude,
calculated using conventional methods.

\subsection{Light-front structure functions}

The two triangle diagrams can be written in the form of a sum of tensors in the
external momenta, multiplied by scalar functions, which we call (covariant)
structure functions. After splitting a covariant diagram in light-front 
time-ordered ones, these can be written again in terms of tensors multiplied by
functions of the external momenta. The latter are called light-front structure
functions. They are not invariant  as they are not defined by four-dimensional
invariant integrals, but rather by three-dimensional integrals. The different
structure functions have different divergences and they must be treated
according to their types of divergence, which we enumerate.

\begin{enumerate}
\item {\em Light-front structure functions without transverse
divergences}.  Neither the covariant nor the light-front formulation
contains any divergences.  Integration over $k^-$ suffices to prove
equivalence. Minus regularization is not allowed.

\item {\em Light-front structure functions with cancelling transverse
divergences. } The individual light-front time-ordered diagrams contain
divergences not present in the covariant amplitude.  Application of
minus regularization to the time-ordered diagrams is not allowed. We
show that the divergences cancel if all the time-ordered diagrams are
added, and that their sum equals the corresponding covariant
amplitude.

\item {\em Light-front structure functions with overall transverse
divergences}.  Divergences appear in the covariant amplitude as well as
in the light-front time-ordered diagrams. We apply BPHZ regularization
to the covariant amplitude and minus regularization to the 
time-ordered diagrams.  

\end{enumerate} 
For the first two cases one can prove equivalence using analytic methods alone.
The proof of equivalence can be found in Refs.~\cite{LB95b,SB98a}. For the
structure functions with overall transverse divergences we have to use
numerical techniques. We show that for the decay of a boson at rest, for both
triangle diagrams, one obtains a rotational invariant amplitude, identical to
the covariant calculation using BPHZ regulariztion.  The fifth diagram
with transverse divergences, the fermion box, will not be discussed.

\subsection{Outline}

The setup for this article is as follows.  In Sec.~\ref{secmr} we
introduce minus regularization.  In Sec.~\ref{secftri} and
Sec.~\ref{secobe} we discuss the equivalence of covariant and light-front
perturbation theory for the fermion triangle and the one-boson exchange
correction. In both cases we start with the covariant calculation and
do the BPHZ regularization if necessary. Then we calculate the
light-front time-ordered diagrams and apply the method mentioned above.
In both cases, we conclude by giving a numerical example of rotational
invariance.

\section{Minus regularization}
\label{secmr}

Minus regularization is inspired by the BPHZ method of regularization,
which gives finite and covariant results. By construction, we ensure
that minus regularization does the same. First we sketch the method
in the case of one-loop diagrams with one independent
external momentum (self-energies),
and next when two independent external momenta (triangle diagrams) are present.
We conclude by generalizing this to a one-loop diagram with $n$
external momenta. For convenience, we shall assume in the latter case
that only logarithmic and
linear divergences are present, such that only the first term of the
Taylor expansion around the renormalization point needs to be subtracted.

Wherever we use the word ``amplitude'' in this section, we refer to an invariant
function of the external momenta. It is understood that the integrals defining
the invariant functions are formally written down in terms of four-dimensional
integrals, which are split into time-ordered pieces by integration over $k^-$.

\subsection{One external momentum}
\label{ssec1}

First we discuss the simple case of one external momentum, which
can be applied for self-energy diagrams.

\subsubsection{BPHZ regularization}
We start with the BPHZ regularization method, which can be applied to
covariant diagrams.
The amplitude has the following form
\begin{eqnarray}
F(q^2) &=& \int {\rm d}^4k\; I_{\rm cov}(q^2;k)\nonumber\\
\label{Fcone}
       &=& F(0) + q^2 F'(0) + \dots
\end{eqnarray}
where $I_{\rm cov}(q^2,k)$ is the covariant integrand generated by
applying standard Feynman rules.
BPHZ regularization renders the amplitude finite by subtracting the infinite parts.
We choose the point $q^2=0$ as the renormalization point, around
which we expand the amplitude in a Taylor series. The higher
orders in the expansion \r{Fcone} are denoted by the elipsis. The regularized
amplitude is then
\begin{equation}
F^{\rm R}(q^2) = F(q^2) - F(0).
\end{equation}
However, this is a purely formal operation, since we are subtracting
two infinite quantities. It is better to write
\begin{eqnarray}
F^{\rm R}(q^2) &=& \int {\rm d}^4k\;
\left( I_{\rm cov}(q^2;k) - I_{\rm cov}(0;k)\right)
\nonumber\\
\label{Fcrone}
&=& \int_0^{q^2}{\rm d} {q'^2} \; \int {\rm d}^4k\;
\frac{\partial}{\partial q'^2} I_{\rm cov}(q'^2;k).
\end{eqnarray}
This guarantees that the amplitude becomes finite.
 
\subsubsection{Minus regularization}
 
Typical for minus regularization is that one writes the amplitude, as well
as the renormalization point, in light-front coordinates.
The covariant choice $q^2=0$ corresponds to $q^- = {{q^\perp}^2}/({2q^+})$.
A time-ordered amplitude corresponding to the covariant form
\r{Fcone} can be written in light-front coordinates as follows
\begin{eqnarray}
F(q^-\!\!,q^+\!\!,q^\perp) &=& \int {\rm d}^3k\; 
I_{\rm lfto}(q^-\!\!,q^+\!\!,q^\perp\!;k)
\label{Ftone}
\\
 = F(\frac{{q^\perp}^2}{2q^+}\!,q^+\!\!,q^\perp) &+&
2q^+(q^-\!\! - \frac{{q^\perp}^2}{2q^+})
F'(\frac{{q^\perp}^2}{2q^+}\!,q^+\!\!,q^\perp) + \dots
\nonumber
\end{eqnarray}
where $I_{\rm lfto}$ is the integrand of the light-front time-ordered
diagram, which was generated by integrating the covariant integrand
$I_{\rm cov}$ over $k^-$ as is explained in Ref.~\cite{LB95b}.
The prime denotes differentiation with respect to $q^-$.
Similar to Eq.~\r{Fcrone} we can write the regularized amplitude as
\begin{equation}
F^{\rm MR}(q^-\!\!,q^+\!\!,q^\perp) =\hspace{-.12cm}
\int_\frac{{q^\perp}^2}{2q^+}^{q^-} {\rm d}q'^- \hspace{-.12cm}
\int {\rm d}^3k\;
\frac{\partial}{\partial q'^-} I_{\rm lfto}(q'^-\!\!,q^+\!\!,q^\perp;k).
\end{equation}
So far we have described the minus regularization method introduced
by Ligterink and Bakker \cite{LB95a}.

\subsection{Two external momenta}
\label{ssec2}
In Ref.~\cite{LB95a} three self-energy diagrams were discussed. For
the triangle diagram the minus regularization technique needs 
to be extended\footnote{We suggest the name MR$^+$.}.   We will tune
the technique by comparing it to BHPZ regularization.
\subsubsection{BPHZ regularization}
The amplitude has the following covariant form
\begin{eqnarray}
F(q_1^2,q_2^2,q_1\!\cdot\!q_2) = \int {\rm d}^4k\;
I_{\rm cov}(q_1^2,q_2^2,q_1\!\cdot\!q_2;k)\nonumber\\
\label{Fctwo}
       = F(\tilde{0}) + q_1^2 F'_{1}(\tilde{0}) + q_2^2 F'_{2}(\tilde{0})
+ q_1\!\cdot\!q_2 F'_{3}(\tilde{0}) + \dots
\end{eqnarray}
where $\tilde{0}$ is the
renormalization point $q_1^2=q_2^2=q_1\!\cdot\!q_2=0$ and $F'_i$ is the 
derivative of $F$ with respect to the $i$th argument.
\begin{equation}
F^{\rm R}(q_1^2,q_2^2,q_1\!\cdot\!q_2) = F(q_1^2,q_2^2,q_1\!\cdot\!q_2)
- F({\tilde{0}}).
\end{equation}
Again, this is a purely formal operation, since we are subtracting
two infinite quantities. We write
\begin{eqnarray}
F^{\rm R}&&(q_1^2,q_2^2,q_1\!\cdot\!q_2) = \int {\rm d}^4k
\nonumber \\
&&\times \left( I_{\rm cov}(q_1^2,q_2^2,q_1\!\cdot\!q_2;k) - 
I_{\rm cov}(\tilde{0};k)\right).
\end{eqnarray}
We cannot, as in the previous section, differentiate with respect to
all external momenta. We would then subtract finite parts from
the Taylor series, containing physical information. This can be
circumvented by introducing a dummy variable $\lambda$, which parametrizes
a straight line in the space of the invariants
between the actual external momenta $q_1^2,q_2^2,q_1\!\cdot q_2$ and
the renormalization point:
\begin{eqnarray}
\label{bphz}
F^{\rm R}(q_1^2,q_2^2,q_1\!\cdot\!q_2) = \int_0^1 {\rm d} \lambda
\int {\rm d}^4k \; \nonumber\\
\times \frac{\partial}{\partial \lambda}
I_{\rm cov}(\lambda q_1^2,\lambda q_2^2, \lambda q_1\!\cdot\!q_2;k)
\end{eqnarray}
We have verified that the $\lambda$-method gives the correct result for the
case where one independent external momentum occurs.
 
\subsubsection{Minus regularization}
 
Again, we write the amplitude in the light-front
time-ordered case as a three-dimensional integral:
\begin{equation}
F(q_i^-\!,q_i^+\!,q_i^\perp) = \int {\rm d}^3k\;
I_{\rm lfto}(q_i^-\!,q_i^+\!,q_i^\perp;k).
\end{equation}
The regularized amplitude is
\begin{equation}
F^{\rm R}(q_i^-\!,q_i^+\!,q_i^\perp)
= F(q_i^-\!,q_i^+\!,q_i^\perp) - F(r_i^-\!,r_i^+\!,r_i^\perp),
\end{equation}
where $r$ defines the renormalization surface. It is a hypersurface
determined by the following conditions:
\begin{eqnarray}
r_1^2 &=& 2 r_1^- r_1^+ - {r_1^\perp}^2 = 0, \nonumber\\
r_2^2 &=& 2 r_2^- r_2^+ - {r_2^\perp}^2 = 0, \\
r_1\!\cdot\!r_2 &=& r_1^- r_2^+ + r_1^+ r_2^- 
           - r_1^\perp\!\cdot r_2^\perp = 0.  \nonumber
\end{eqnarray}
This set of equations is equivalent to
\begin{equation}
r_1^2 = 0, \;\;\;
r_2  = \chi r_1. \\
\end{equation}
The $r_i^+$ enter in the integration boundaries; therefore we would like them to
remain unaffected by regularization ($r_i^+ = q_i^+$). This implies that
$\chi$ can be found from 
\begin{equation}
\chi= \frac{q_2^+}{q_1^+}.
\end{equation}
The only freedom that remains is the choice for $r_1^\perp$.
Two choices come
easily to mind:  $r_1^\perp=0$ (method MR0) and $r_1^\perp=q_1^\perp$
(method MR1).
\begin{eqnarray}
({\rm MR0}) \hspace{2cm} r_1^\perp &=& 0^\perp
\Rightarrow r_2^\perp = 0^\perp,\\
({\rm MR1}) \hspace{2cm} r_1^\perp &=& q_1^\perp \Rightarrow r_2^\perp =
\chi {q_1^\perp}.
\end{eqnarray}

{TABLE I. The light-front parametrization of the renormalization point $r^\mu$ 
for two equivalent choices of minus regularization, MR0 and MR1.}
\label{tab1}

\begin{tabular}{|c|c|c|}
\hline
&MR0&MR1\\
\hline
$\;(r_1^-,r_1^+,r_1^\perp)\;$ &
$\;\phantom{\chi}(0,q_1^+,0^\perp)\;$ &
$\;\phantom{\chi}(
{{q_1^\perp}^2}/({2q_1^+}), q_1^+,
q_1^\perp)\;$ \\
\hline
$\;(r_2^-,r_2^+,r_2^\perp)\;$ &
$\;\chi \;(0,q_1^+,0^\perp)\;$ &
$\;\chi \;( {{q_1^\perp}^2}/({2q_1^+}), q_1^+, q_1^\perp)\;$ \\
\hline
\end{tabular}
\\
\\

The details are worked out in Table~I. 
The light-front coordinates of the renormalization point are used in
the following way to find the regularized light-front amplitude:
\begin{eqnarray}
F^{\rm MR}&&(q_i^-,q_i^+,q_i^\perp) = \int_0^1 {\rm d} \lambda
\int {\rm d}^3k \;
\frac{\partial}{\partial \lambda}
\nonumber\\&&
\label{mr} \times
I_{\rm lfto}(\lambda (q_i^-\!\!-r_i^-)\! +\! r_i^-,
q_i^+\!,
\lambda (q_i^\perp\!\!-r_i^\perp)\! +\! r_i^\perp;k).
\end{eqnarray}
In this formula we recognize our choice $r_i^+ = q_i^+$.

\subsection{Several external momenta}
\label{ssecn}
 
The method just described can be generalized to the case of a loop with
an arbitrary number of external lines.
The procedure is almost the same as for two external momenta. 
The renormalization surface is given by
\begin{eqnarray}
\label{ncond1}
 r^2_i & = & 2 r^-_i r^+_i - r^{\perp 2}_i = 0, \\
\label{ncond2}
r_i\cdot r_j&  = & r_i^- r_j^+ + r_i^+ r_j^- - r_i^\perp\!\cdot r_j^\perp = 0
\;\;\;(i \not= j).
\end{eqnarray}
These equations are equivalent to
\begin{equation}
r^2_1 = 0, \;\;\;r_i = \chi_{\scriptscriptstyle i} r_1.
\end{equation}
Again, we make the choice to leave the plus components of
the momenta unaffected by regularization: $r^+_i = q^+_i$.
This implies that the $\chi_{\scriptscriptstyle i}$ are
fractional longitudinal light-front momenta.
\begin{equation}
\chi_{\scriptscriptstyle i} = \frac{q_i^+}{q_1^+}.
\end{equation}
Two choices for $r_1^\perp$ are listed below. This
then determines all other $r_i^\perp$.
\begin{eqnarray}
({\rm MR0}) \hspace{2cm} r_1^\perp &=& 0^\perp
\Rightarrow r_i^\perp=0^\perp,\\
({\rm MR1}) \hspace{2cm} r_1^\perp &=& q_1^\perp \Rightarrow r_i^\perp =
\chi_{\scriptscriptstyle i} {q_1^\perp}.
\end{eqnarray}
 
\subsection{Summary}

The way we setup minus regularization does not rely on the structure
of the covariant or the time-ordered diagrams, but works on the level
of the external momenta only.  If an amplitude has a covariant
structure before regularization, minus regularization guarantees that
it remains covariant.  In our implementation of BPHZ regularization, the
renormalization point corresponds to all invariants connected to the
external momenta being equal to zero. These conditions allow minus
regularization to take on a number of forms.  Of these, we shall apply
MR0 and MR1. The main difference between them is that MR0 does not
choose one of the momenta as a preferred direction, and therefore it
explicitly maintains all symmetries of the external momenta.
Furthermore, MR0 gives rise to shorter formulas for the regularized
integrands.

In the next two sections both  methods are being applied to
the parts of two light-front time-ordered triangle diagrams
in the Yukawa model containing transverse divergences, viz the fermion
triangle and the one-boson exchange correction.
 
\section{Equivalence for the fermion triangle}
 
\label{secftri}
In the Yukawa model there is an effective three boson interaction,
because for a fermion loop with a scalar coupling Furry's theorem does not
apply. The leading order contribution to this process is the fermion 
triangle. A scalar boson of mass $\mu$ and momentum $p$ comes in and
decays into two bosons of momentum $q_1$ and $q_2$ respectively.
The fermions in the triangle have mass $m$. The covariant expression for the
amplitude is
\begin{equation}
\label{ftri1}
\ftrix = \int_{\rm Min}
\frac{{\rm d}^4k \;\; {\rm Tr}
\left [(\slash{k}_1 + m)(\slash{k}_2 + m) (\slash{k} + m)\right]} 
{(k_1^2-m^2) (k_2^2  -  m^2) (k^2  -  m^2)} .
\end{equation}
The subscript ``Min'' denotes that the integration is over Minkowski space.
The usual imaginary parts of the Feynman propagators have been dropped.
We have omitted numerical factors and have set the coupling constant to unity.
The momenta $k_1$ and $k_2$ indicated in the diagram are given by
\begin{equation}
k_1 = k - q_1 , \;\;\; k_2 = k + q_2 .
\end{equation}
Of course, by momentum conservation we have
\begin{equation}
p = q_1 + q_2.
\end{equation}
We evaluate the integral \r{ftri1} first in the usual covariant way,
and subsequently carry out $k^-$-integration to produce the light-front time-ordered
diagrams.
Note that integral \r{ftri1} is an ill-defined formula. In
both methods mentioned we have to define what we mean by this integral.

\subsection{Covariant calculation}
The following method is usually applied to calculate the fermion triangle
in a covariant way.
First, one introduces Feynman parameters $x_1$ and $x_2$, and then one
shifts the loop variable $k$ to complete the squares in the denominator.
The result is
\begin{eqnarray}
\label{ftri2} \hspace{-2cm}
\ftri&=&8 \int_0^1 {\rm d}x_1
\int_0^{1-x_1} {\rm d}x_2 \int_{\rm Min} {\rm d}^4k
\\
&\times&
\frac{ m^3 + m \left( 3{k}^2 + {\cal P}^2 \right) + {\rm terms\; odd\;in\;}k}
{\left( k^2 - m^2 + {\cal Q}^2 \right)^3},
\nonumber\end{eqnarray}
with
\begin{eqnarray}
{\cal Q}^2 &=&x_1(1\!-\!x_1)\; q_1^2 + x_2(1\!-\!x_2)\; 
q_2^2 + 2 x_1 x_2\;q_1\!\cdot\!q_2,\\
{\cal P}^2 &=& x_1 (3 x_1\!-\!2) q_1^2 + x_2 (3 x_2\! -\! 2 ) q_2^2
\nonumber\\
&&+ \left( 2 (x_1\! +\! x_2) - 6 x_1 x_2 - 1 \right) q_1\!\cdot\!q_2.
\end{eqnarray}
As a last step, we remove the terms odd in $k$.

\subsection{BPHZ regularization}
The regularized fermion triangle can be found by
applying the BPHZ regularization scheme \r{bphz} to the covariant 
formula \r{ftri2}.
The integral is now finite; so we can do the Wick rotation and
perform the $k$ integrations.
The result is
\begin{eqnarray}
\ftri^{{\rm \; R}} =
 {-4 \pi^2 i} \!\int_0^1\!\!\! \!{\rm d}x_1\!\!
\int_0^{1-x_1}\!\!\!\!\!\!{\rm d}x_2\!
\int_0^1 {\rm d}\lambda\nonumber\\
\times
\frac{ m \left(m^2 ( 5 {\cal Q}^2 - {\cal P}^2) - 6 \lambda {\cal Q}^4\right)}
{\left(m^2 - \lambda {\cal Q}^2 \right)^2}.
\end{eqnarray}
The superscript R indicates an integral regularized according to the
BPHZ method.

\subsection{Light-front calculation}
Using the method given in Ref.~\cite{LB95b} we proceed as follows.
The $k^-$ dependence of a spin projection in the numerator is removed by 
separating it into an on-shell spin projection and an instantaneous part:
\begin{equation}
\slash{k}_i + m =(\slash{k}_{i\; \rm on}+m)+(k^- - k^-_{i\; \rm on})\gamma^+,
\end{equation}
where the vector $k^\mu_{i\; \rm on}$ is given by
\begin{equation}
\left(k^-_i,k^+_i,k^\perp_i\right)_{\rm on}=
\left(\frac{{k^\perp_i}^2 + m^2}{2 k^+_i},k^+_i,{k^\perp_i} \right).
\end{equation}
Factors like $(k^-\! - k^-_{i\; \rm on})$ can be divided out against
propagators and this cancellation gives rise to instantaneous fermions.
The integration over $k^-$ is performed by contour 
integration. The poles of the propagators are given by
\begin{eqnarray}
\label{pole1a}
H^-    &=& \frac{{k^\perp}^2 + m^2 }{2k^+} , \\
\label{pole2}
H^-_1  &=& q^-_1-\frac{{k_1^\perp}^2 + m^2 }{2k^+_1} , \\
\label{pole3}
H^-_2  &=& -q^-_2+\frac{{k_2^\perp}^2 + m^2 }{2k^+_2} .
\end{eqnarray}
This integration gives rise to the different time-ordered diagrams, as 
explained in more detail in Refs.~\cite{LB95b,SB98a}. The result is
\begin{eqnarray}
\ftri &=& \ftriapp + \ftriaip + \ftriapi \nonumber\\
\vspace{.2cm}
      &+& \ftribpp + \ftribip + \ftribpi
\end{eqnarray}
The diagrams on the right-hand side are light-front time-ordered diagrams.
Time goes from left to right. The pictures can be recognized as
time-ordered diagrams because of the time-ordering of the
vertices and the occurrence of instantaneous fermions, indicated by
a horizontal tag. Explicitly;
\begin{eqnarray}
\ftriapp &=& 2 \pi i \int {\rm d}^2k^\perp \int_0^{q_1^+}
\frac{ {\rm d}k^+ }
{8 k_1^+ \! k_2^+ {k}^+}
\nonumber\\ &\times& \; 
\label{ftriapp}
\frac{{\rm Tr}\left[(\slash{k}_{1\rm on} + m) (\slash{k}_{2\rm on} + m) 
(\slash{k}_{\rm on} + m)\right] }
{(H_1^-\! \!-\! H_2^-)(H_1^-\! \!-\! H^-)},
\\ \nonumber \\ \nonumber \\
\vspace{.5cm}
\ftriaip &=&  2 \pi i \int {\rm d}^2k^\perp \int_0^{q^+_1}
\frac{ {\rm d}k^+ }
{8 k_1^+ \! k_2^+ {k}^+}
\nonumber\\ &\times&
\label{ftriaip}
{}
\frac{{\rm Tr}\left[(\slash{k}_{1\rm on} + m) \gamma^+
(\slash{k}_{\rm on} + m)\right]} {H_1^-\! \!-\! H^-} ,
\\ \nonumber \\ \nonumber \\  
\ftriapi &=&  2 \pi i \int {\rm d}^2k^\perp \int_0^{q^+_1}
\frac{ {\rm d}k^+ }
{8 k_1^+ \! k_2^+ {k}^+}
\nonumber\\ &\times&
\label{ftriapi}
{}
\frac{{\rm Tr}\left[(\slash{k}_{1\rm on} + m) (\slash{k}_{2\rm on} + m)
\gamma^+\right]} {H_1^-\! \!-\! H_2^-},
\\ \nonumber \\ \nonumber \\  
\ftribpp &=&-2 \pi i \int {\rm d}^2k^\perp \int_{-q_2^+}^0
\frac{ {\rm d}k^+ }
{8 k_1^+ \! k_2^+ {k}^+}
\nonumber\\ &\times& \;
\label{ftribpp}
\frac{{\rm Tr}\left[(\slash{k}_{1\rm on} + m) (\slash{k}_{2\rm on} + m)
(\slash{k}_{\rm on} + m)\right] }
{(H_1^-\! \!-\! H_2^-)(H^-\! \!-\! H_2^-)},
\\ \nonumber \\ \nonumber \\  
\ftribip &=&  -2 \pi i \int {\rm d}^2k^\perp \int_{-q_2^+}^0
\frac{ {\rm d}k^+ }
{8 k_1^+ \! k_2^+ {k}^+}
\nonumber\\ &\times&
\label{ftribip}
\frac{{\rm Tr}\left[\gamma^+ (\slash{k}_{2\rm on} + m) 
(\slash{k}_{\rm on} + m)\right]} {H^-\! \!-\! H_2^-} ,
\\ \nonumber \\ \nonumber \\  
\ftribpi &=&  -2 \pi i \int {\rm d}^2k^\perp \int_{-q_2^+}^0
\frac{ {\rm d}k^+ }
{8 k_1^+ \! k_2^+ {k}^+}
\nonumber\\ &\times&
\label{ftribpi}
\frac{{\rm Tr}\left[(\slash{k}_{1\rm on} + m) (\slash{k}_{2\rm on} + m)
\gamma^+\right]} {H_1^-\! \!-\! H_2^-}.
\end{eqnarray}
Note that the diagrams \r{ftriapi} and \r{ftribpi}
with the instantaneous exchanged fermions have
the same integrand. However, the longitudinal momentum 
$k^+$ has a different sign. 

Although we could have expected diagrams with two instantaneous
fermions, we see that they are not present. This is so because we use a
scalar coupling and therefore two $\gamma^+$ matrices becoming
neighbors give~$0$.  No so-called forced instantaneous loops 
are present.  These FILs obscure the equivalence of light-front and
covariant perturbation theory and have been analyzed in Ref.~\cite{SB98a}.
They will not be discussed in this paper, since they are related to
longitudinal divergences.

The traces can be calculated. We obtain
\begin{eqnarray}
&{\rm Tr}&\left[(\slash{k}_{1\rm on} + m) (\slash{k}_{2\rm on} + m)
(\slash{k}_{\rm on} + m)\right]\nonumber\\
&&= 4 m (m^2 
+ {k}_{1\rm on}\!\cdot\!{k}_{\rm on} + {k}_{2\rm on}\!\cdot\!{k}_{\rm on}
+ {k}_{1\rm on}\!\cdot\!{k}_{2\rm on}),
\\
&{\rm Tr}&\left[(\slash{k}_{1\rm on} + m) (\slash{k}_{2\rm on} + m)
\gamma^+\right] = 4 m \left( 2 k^+\!\!- q_1^+\!\! 
                   + q_2^+ \right)\!,\hspace{-.3cm}
\\
&{\rm Tr}&\left[(\slash{k}_{1\rm on} + m) \gamma^+
(\slash{k}_{\rm on} + m)\right]\; = 4 m \left( 2 k^+\! - q_1^+ \right),
\\
&{\rm Tr}&\left[\gamma^+ (\slash{k}_{2\rm on} + m)
(\slash{k}_{\rm on} + m)\right]\; =  4 m \left( 2 k^+\! + q_2^+ \right).
\end{eqnarray}
We see that the high orders in $k^\perp$ have disappeared in the traces. 
However, logarithmic divergences remain in all light-front time-ordered
diagrams \r{ftriapp}-\r{ftribpi}. We tackle them with minus 
regularization, as introduced in the previous subsection.

\subsection{Equivalence}
 
As the fermion triangle is a scalar amplitude,
there is only one structure function present. It belongs to the
first category we mentioned in the Introduction: it is logarithmically 
divergent, but has no longitudinal divergences.

\subsubsection{Light-front structure functions with transverse divergences} 

We applied minus regularization to the integrands of the six
light-front time-ordered diagrams,  using both the MR0 and MR1 methods.
We used {\sc mathematica} to do the substitution and the differentiation with
respect to $\lambda$, given by Eq.~\r{mr}. 
However, {\sc mathematica} was not able to do the
integration, neither analytically nor numerically. Therefore the
integrand was implemented in {\sc fortran} which was well capable of doing
the four-dimensional integration using {\sc imsl} routines based on Gaussian
integration.
 
\begin{figure}
\hspace{2.5cm} \epsfxsize=4cm \epsffile[ 0 0 205 227]{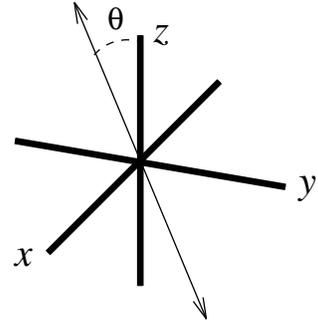} 
\vspace{.2cm}
\caption{A boson is at rest and decays into two particles flying off
\label{figtheta}
in opposite directions. The angle $\theta$ is the angle between the momentum of
one of the fermions and the $z$-axis. }
\end{figure}

Because  the integrations cannot be done exactly, we saw no
possibility of giving a rigorous proof of the equivalence of
light-front and covariant perturbation theory. Instead we make a
choice for the parameters, such as the masses and the external momenta,
and show that our method gives the same result as the
covariant calculation with BPHZ regularization.  We calculated the
decay amplitude of a scalar boson at rest, as is pictured in
Fig.~\ref{figtheta}. 

From a physical point of view, there is no preferred direction, and
therefore we demand that our choice of the coordinates of the
light-front have no influence on the outcome of the calculation. The
decay amplitude, which is a scalar quantity, should give the same
result for each possible direction in which the bosons can fly off.

\begin{figure} 
\epsfxsize=8.7cm \epsffile[75 480 505 795 ]{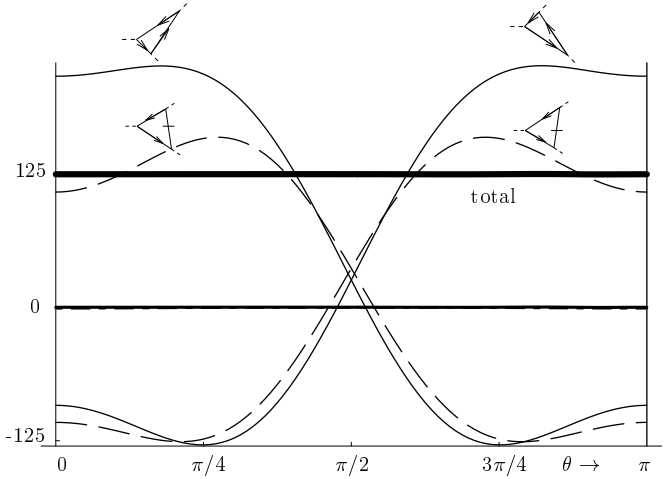} 
\vspace{.2cm}
\caption{The thick line at a value of 125 represents the sum
of the six light-front time-ordered amplitudes. It is independent of the
angle $\theta$, defined in the previous figure.
The four largest contributions come from the diagrams without instantaneous 
parts (solid lines) and the diagrams with an instantaneous exchanged
\label{figdrie} 
fermion (dashed lines), as indicated by the diagrams.} 
\end{figure}

\begin{figure}
\epsfxsize=8.8cm \epsffile[75 480 500 750 ]{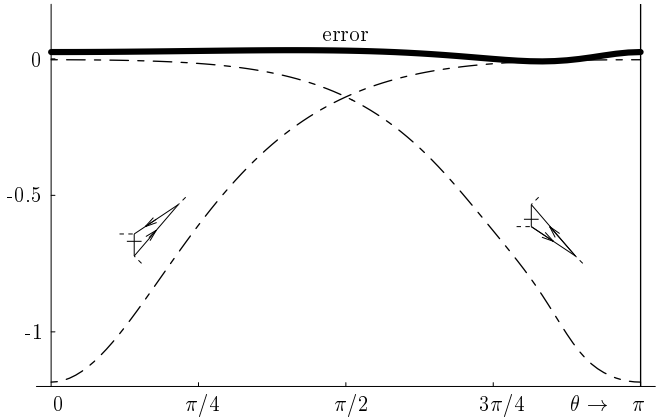}
\vspace{.2cm}
\caption{The amplitudes of the two small contributions (double-dashed lines) and
the difference between the sum of the six light-front time-ordered diagrams
\label{figvier}
and the covariant amplitude (thick solid line).}
\end{figure}

\begin{figure}
{\epsfxsize=8.7cm \epsffile[190 420 430 780]{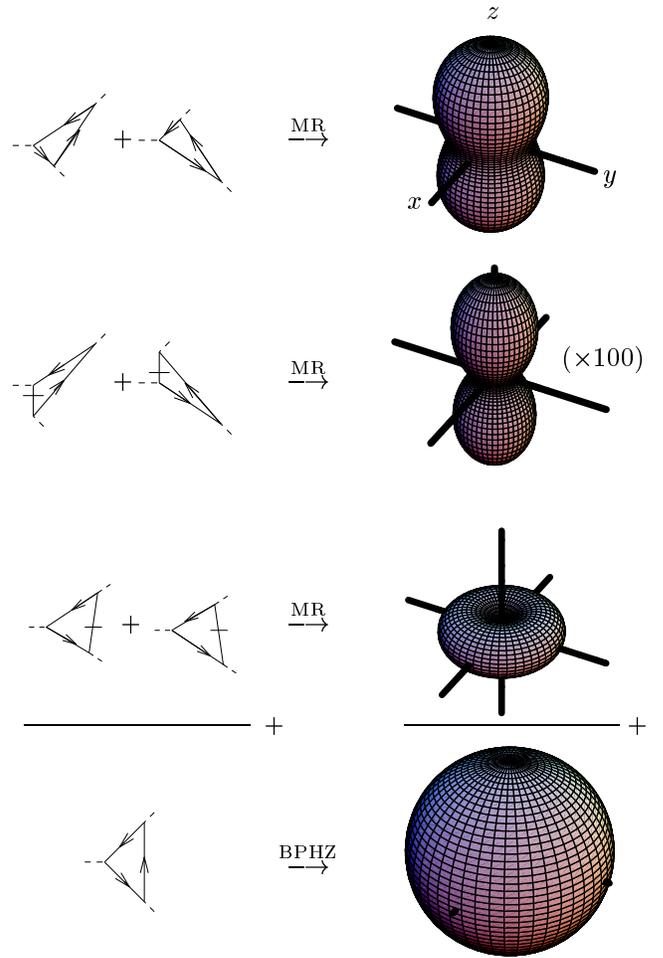}}
\vspace{.2cm}
\caption{Commutative diagram of the boson decay amplitude.
The boson is at rest in the origin and decays.
The outgoing bosons fly off in opposite directions.
Points on the surfaces have polar coordinates ($A, \theta, \phi)$,
where $A$ the is magnitude of the amplitude and $\theta$ and $\phi$ are
the polar angles of the momentum of one of the outgoing particles, as
defined in Fig.~\ref{figtheta}.  Because the diagrams on
the second line are very small, the scale  has been enlarged by a factor of 100.
For the light-front time-ordered diagrams on the first three lines 
minus regularization (both MR0 and MR1) is used, 
\label{figvijf}
for the covariant diagram on the last line we used BPHZ regularization.}
\end{figure}

There are six minus-regularized light-front
time-ordered fermion triangle diagrams contributing to the boson decay.
Each individual light-front time-ordered diagram has a manifest
rotational invariance in the $x$-$y$-plane, and therefore we expect the
same for the sum.  However, since light-front perturbation theory
discriminates between the $z$-direction and the other space-like
directions, the light-front time-ordered diagrams can (and should)
differ as a function of the angle, $\theta$, between the momentum of one
of the particles flying off and the $z$-axis. The absolute value of
the momentum was fixed.
It is not immediately clear that the sum should be invariant.  This
investigation becomes more interesting since it is believed \cite{BL91}
that rotational invariance is broken in naive light-cone quantization
of the Yukawa model.  However, the results shown in
\mbox{Figs.~\ref{figdrie}-\ref{figvijf}} demonstrate that rotational
invariance is not broken.  Note that we have dropped the factor~$-i$
common to all diagrams.

Two light-front time-ordered diagrams \r{ftriaip},\r{ftribip}
contributing to the boson decay and indicated by double-dashed lines
are so small they can hardly be identified in Fig.~\ref{figdrie}. 
In Fig.~\ref{figvier} we depict these two on a scale that is a factor
of 100 larger. In the same figure we show 
 the difference of the sum of the six light-front time-ordered diagrams 
(using MR1 and 128 points in every integration variable)
and the covariant result. It has a maximum of 0.03\%.
 
In Figs.~\ref{figdrie} and~\ref{figvier} we see that interchanging the outgoing
bosons is the same as replacing  $\theta$ by $\pi - \theta$.

We verified that the individual diagrams are rotational invariant
around the $z$-axis. We illustrate this in Fig~\ref{figvijf}.

Summing up, we find that the sum of the minus regularized 
light-front time-ordered diagrams is rotational invariant.
The deviation from the covariant result is smaller than 0.03\%. 
It is illustrated in Fig.~\ref{figvier}. We checked, by varying the number
of integration points, that the deviations are due to numerical
inaccuracies only. We conclude that, for the fermion 
triangle, the covariant calculation in combination with the BPHZ regularization
scheme gives the same result as the light-front calculation in combination
with minus regularization.

\section{Equivalence for the one-boson exchange correction}
\label{secobe}
The second process under investigation was studied before by
Burkardt and Langnau \cite{BL91}. 
A scalar boson of mass $\mu$ and momentum $p$ decays into two
fermions of mass $m$ and momentum $q_1$ and $q_2$ respectively.
The lowest order correction to this process is the
one-boson exchange correction. The amplitude is given by the integral
\begin{equation}
\label{obe1}
\obex   = \int_{\rm Min}
\frac{{\rm d}^4k \;\; (\slash{k}_1 + m)(\slash{k}_2 + m)}
{(k_1^2-m^2) (k_2^2  -  m^2)
(k^2  -  \mu^2)} .
\end{equation}
Again, this equation is undefined as it stands. First we have to make
it a well-defined object. In Sec.~\ref{1207} we apply the 
covariant method and in Sec.~\ref{1208} we use light-front coordinates.
\subsection{Covariant calculation}
\label{1207}
Using Feynman parametrization the one-boson exchange correction
can be rewritten as
\begin{equation}
\label{obe2} 
\obe=2 \int_0^1 {\rm d}x_1
\int_0^{1-x_1} {\rm d}x_2 \int_{\rm Min} {\rm d}^4k 
\end{equation}
\vspace{-.5cm}
\begin{eqnarray*}
\frac{ \slash{k}^2 +
\left( (1\!\!-\!\!x_1) \slash{q}_1\! +\!\!x_2) \slash{q}_2\! +\! m \right)
\left( -\!x_1 \slash{q}_1\! -\! (1\!\!-\!\!x_2) \slash{q}_2 \!+\! m \right)}
{\left( k^2 - {\cal M}^2 + {\cal Q}^2 \right)^3}\\
+ { \rm \; terms \; odd \; in \;}k 
\end{eqnarray*}
with
\begin{eqnarray}
{\cal M}^2 &=&(x_1 + x_2) m^2 + (1-x_1-x_2) \mu^2,\\
{\cal Q}^2 &=& x_1(1-x_1) q_1^2 + x_2(1-x_2) q_2^2 + 2 x_1 x_2q_1\!\cdot\!q_2,
\end{eqnarray}
and where terms odd in $k$ in the numerator are not specified, since they
will be removed according to the covariant prescription. We also define
\begin{equation}
{\cal P}^2 = {\cal Q}^2 + (1 - x_1 - x_2) q_1\! \cdot\! q_2.
\end{equation}

>From Eq.~\r{obe2} we can infer that the Dirac structure of the diagram is
\begin{equation}
\label{xyz5}
\vspace{-.2cm}
\obe= F^1 + F^{2\mu} \gamma_\mu
+ F^{3\mu \nu} \frac{1}{2} \left[ \gamma_\mu, \gamma_\nu \right].
\vspace{.2cm}
\end{equation}
where the vector part contains a symmetric and an anti-symmetric part,
\begin{equation}
F^{2\mu} = F^{2\rm s}(q_1^{\mu} + q_2^{\mu})
        + F^{2\rm a}(q_{1}^{\mu} - q_{2}^{\mu}),
\end{equation}
and the tensor part has the form
\begin{equation}
F^{3\mu \nu} = (q_{1}^{\mu} q_{2}^{\nu} - q_{1}^{\nu} q_{2}^{\mu}) F^3.
\end{equation}
The functions $F^i$ depend on the masses and the
external momenta $q_1^2$, $q_2^2$ and $q_1\!\cdot\!q_2$. 
If we define the integral operator
\begin{equation}
\label{xyz1}
I[f] =2 \int_0^1 {\rm d}x_1 \int_0^{1-x_1} {\rm d}x_2 \int_{\rm Min}{\rm d}^4k
\left( k^2 -{\cal M}^2\!+ {\cal Q}^2 \right)^{-3}\!\! f,
\end{equation}
then we have, using $\slash{q}_1^2 = q_1^2$, etc.
\begin{eqnarray}
\label{F1eq}
F^1 &=& I \left[ k^2+ m^2- {\cal P}^2\right],\\
F^{2\rm a} &=& 2 m I \left[ 1 - x_1 - x_2 \right], \\
F^{2\rm s} &=& 2 m I \left[ - x_1 + x_2 \right], \\
F^3&=& I \left[ 1 - x_1 - x_2\right] .
\end{eqnarray}
We see that the only function which needs to be regularized is $F^1$.
The functions $F^2$ and $F^3$ are convergent and do not require
regularization in a covariant calculation.
\subsection{BPHZ regularization}
The regularized structure function $F^{1 \rm R}$ can be found by 
applying the BPHZ regularization scheme \r{bphz} to the structure
function~\r{F1eq}.
The integral is now finite; so we can do the Wick rotation and
perform the $k$ integrations.
\begin{eqnarray}
F^{1{\rm R}}(q_1^2,q_2^2,q_1\! \cdot \!q_2) =
-2 \pi^2 i \!\int_0^1\!\!\! \!{\rm d}x_1\!\!
\int_0^{1-x_1}\!\!\!\!\!\!{\rm d}x_2\!
\int_0^1 {\rm d}\lambda\nonumber\\
\times \left( \frac{{\cal Q}^2 (\lambda {\cal P}^2 - m^2)}
            {2 ({\cal M}^2 - \lambda {\cal Q}^2)^2} +
       \frac{{\cal Q}^2 + \frac{1}{2}{\cal P}^2}
            {{\cal M}^2 - \lambda {\cal Q}^2} \right).
\end{eqnarray}
We have not been able to do all three integrations exactly.  The
$\lambda$ integration and one of the $x$ integrations can be done
analytically, and the remaining integration numerically.  As $F^{2\mu}$
and $F^3$ do not need to be regularized, this concludes the covariant
calculation of the one-boson exchange correction.

\subsection{Light-front calculation}
\label{1208}
In our previous paper \cite{SB98a} it was shown how to derive the
light-front time-ordered diagrams corresponding to the covariant 
diagram \r{obe1} using $k^-$-integration.  One can write the
time-ordered diagrams individually, or one can combine propagating and
instantaneous parts into so-called blinks. Blinks, introduced by
Ligterink and Bakker \cite{LB95b}, have the advantage that the
$1/k^+$-singularities cancel and the number of diagrams is reduced.

In the two triangle diagrams studied here it makes no difference
whether blinks are used or not. In the case of the fermion
triangle we calculated light-front time-ordered diagrams.
Here we use blinks, to demonstrate that our technique also works in
this case.  The one-boson exchange correction has two blinks:

\vspace{-.2cm}
\begin{equation}
\obe = \obebb + \obeba
{}
\end{equation}

\vspace{ .4cm} 
The poles of the two fermion propagators in the triangle are 
given by Eqs.~\r{pole2} and
\r{pole3}. The pole of the boson propagator is given by
\begin{equation}
\label{pole1b}
H^-    = \frac{{k^\perp}^2 + \mu^2 }{2k^+} . \\
\end{equation}
The amplitudes including blinks are 
\begin{eqnarray}
\label{obebb}
{}
\obebb =
- 2 \pi i \int {\rm d}^2k^\perp \int_{-q_2^+}^0
\frac{ {\rm d}k^+ }
{8 k_1^+ \! k_2^+ {k}^+}\nonumber\\
\times \; \frac{(\slash{k}_{2\rm on} - \slash{p} + m)
(\slash{k}_{2\rm on} + m) }
{(H_1^-\! \!-\! H_2^-)(H^-\! \!-\! H_2^-)} ,
\\
\label{obeba}
{}
\obeba =
2 \pi i \int {\rm d}^2k^\perp \int_0^{q_1^+}
\frac{ {\rm d}k^+ }
{8 k_1^+ \! k_2^+ {k}^+}\nonumber\\
\times \; \frac{(\slash{k}_{1\rm on} + m)
(\slash{k}_{1\rm on} + \slash{p} + m) }
{(H_1^-\! \!-\! H_2^-)(H_1^-\! \!-\! H^-)} .
\end{eqnarray}
We will now focus on the blink in Eq.~\r{obeba}.
It  simplifies  because we can use
\begin{equation}
\slash{k}_{1\rm on} \slash{k}_{1\rm on} = {k}_{1\rm on} \cdot {k}_{1\rm on}
= m^2.
\end{equation}
Therefore we obtain
\begin{eqnarray}
\label{obeba2}
\obeba =
2 \pi i \int {\rm d}^2k^\perp \int_0^{q_1^+}
\frac{ {\rm d}k^+ }
{8 k_1^+ \! k_2^+ {k}^+}\nonumber\\
\times \; \frac{2 m^2 + \slash{k}_{1\rm on} (\slash{p} + 2 m)}
{(H_1^-\! \!-\! H_2^-)(H_1^-\! \!-\! H^-)} .
\end{eqnarray}
In the same way as we did for the covariant amplitude we can identify
the different Dirac structures

\begin{eqnarray}
\label{obeba3}
\vspace{-.4cm}
\obeba =
F^1_1 + F^{2\mu}_{1} \gamma_\mu
+ F^{3\mu\nu}_{1} \frac{1}{2} \left[ \gamma_\mu, \gamma_\nu \right].
\\ \nonumber
\end{eqnarray}

Although at first sight it looks as if the diagram in Eq.~\r{obeba2}
has a covariant structure, covariance is spoiled by the integration
boundaries for $k^+$.  Therefore these functions are not covariant
objects.  We have to investigate equivalence for the structure
functions separately.

The light-front structure function $F^1_1$ can be  
found by taking the trace of Eq.~\r{obeba2}, since
all the other structures are traceless. Carrying out the traces one finds
\begin{eqnarray}
\label{F11}
F^1_1 =
  2 \pi i \int {\rm d}^2k^\perp \int_0^{q_1^+}
\frac{ {\rm d}k^+ }
{8 k_1^+ \! k_2^+ {k}^+}\nonumber\\
\times \; \frac{2 m^2 + {k}_{1\rm on}\!\cdot\!p}
{(H_1^-\! \!-\! H_2^-)(H_1^-\! \!-\! H^-)} .
\end{eqnarray}
The other structures of the blink diagram \r{obeba2} are
\begin{eqnarray}
\label{F21}
F_{1}^{2\mu} =
  2 \pi i \int {\rm d}^2k^\perp \int_0^{q_1^+}
\frac{ {\rm d}k^+ }
{8 k_1^+ \! k_2^+ {k}^+}\nonumber\\
\times \; \frac{2 m \; ({k}_{1\rm on})^\mu}
{(H_1^-\! \!-\! H_2^-)(H_1^-\! \!-\! H^-)} .
\end{eqnarray}

\begin{eqnarray}
\label{F31}
F_{1}^{3\mu\nu} =
  2 \pi i \int {\rm d}^2k^\perp \int_0^{q_1^+}
\frac{ {\rm d}k^+ }
{8 k_1^+ \! k_2^+ {k}^+}\nonumber\\
\times \; \frac{({k}_{1\rm on})^\mu \; p^\nu}
{(H_1^-\! \!-\! H_2^-)(H_1^-\! \!-\! H^-)} .
\end{eqnarray}

In a similar way we can derive the structure functions corresponding
to the other blink diagram. 

\subsection{Equivalence}

We can identify the different types of divergences, as explained
in the Introduction.

\subsubsection{Light-front structure functions without transverse
divergences}
The parts of the blinks without any ultraviolet divergences
are $F^{2\mu}_{i}$ and $F^{3\mu\nu}_{i}$, except for $\mu$ being $-$.
No cancellations need to be found and no regularization is necessary.

\subsubsection{Light-front structure functions with cancelling
transverse divergences}
In the last two structure functions we
see something odd happening. Both $F_{i}^{2\mu}$ and $F_{i}^{3\mu\nu}$
are divergent for $\mu$ being $-$. However, these divergences
are not present in the covariant structure functions $F^{2\mu}$
and $F^{3\mu\nu}$. It would we illegal to apply minus regularization,
since the covariant amplitude does not need to be regularized. 
We found that the divergences corresponding to the first blink
cancel exactly against those of the second blink. 
To simplify the calculation
we use internal variables $x'$ and $k^\perp$ 
and external variables $\chi$, $q_i^-$ and $q_i^\perp$. These
are introduced in Appendix~\ref{app1}. 

We have to verify the following relation of equivalence
\begin{equation}
F^{2-} = F^{2-}_1 + F^{2-}_2.
\end{equation}
According to the reasons mentioned above we have to
demand that the divergent parts in the right-hand side cancel.
We find that only the highest order contribution in $k^\perp$
contributes to a divergent integral, because we can write
\begin{eqnarray}
F^{2-}_i = \int {\rm d}^2k^\perp \left( \frac{f^{2-}_i}{{k^\perp}^2}
  + g^{2-}_i(k^\perp) \right),
\end{eqnarray}
where $g^{2-}_i(k^\perp)$ is the part of the integrand without 
ultraviolet divergences, and the term with $f^{2-}_i$ gives rise
to a logarithmically divergent integral. We have to check if we have
\begin{equation}
\label{cancel}
f^{2-}_1 + f^{2-}_2 = 0.
\end{equation}
In Appendix~\ref{app1} the full formulas for the functions $f^{2-}_i$ are given,
from which it follows that condition~\r{cancel} holds. 
For $\mu$ being $-$ in the
structure function $F^{3\mu\nu}_{1}$ one can apply the same method.

\subsubsection{Light-front structure functions with transverse
divergences}

\begin{figure}
{\epsfxsize=8.7cm \epsffile[200 500 420 790]{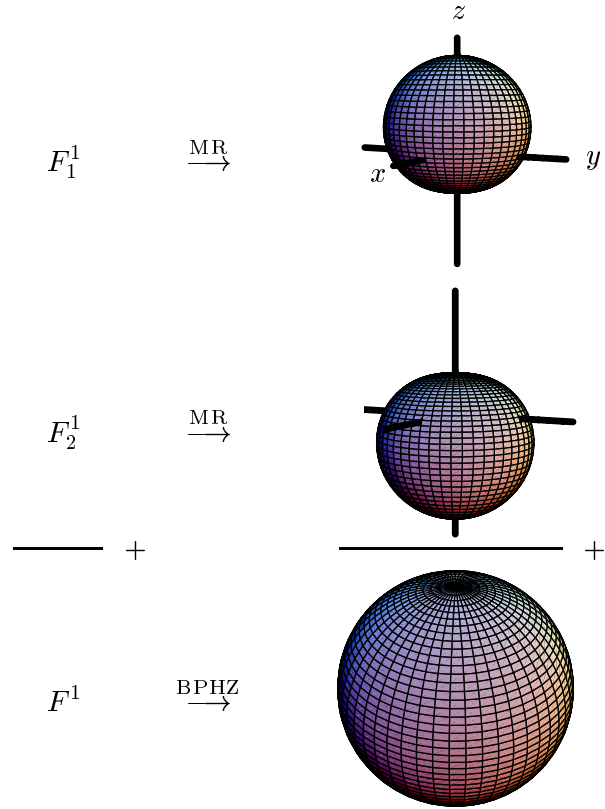}}
\vspace{.2cm}
\caption{Commutative diagram of the one-boson exchange correction.
A boson decays at rest.
The outgoing fermions fly off in opposite directions.  The distance from
the origin gives the amplitude of the regularized diagram
for the fermion flying off in this direction. 
For the light-front structure functions on the first two lines,
minus regularization (MR1) is used; for the covariant structure 
\label{figzes}
function on the last line, we used BPHZ regularization.}
\end{figure}

The structure function $F^1$ in the covariant
calculation contains an ultraviolet divergence. In the light-front
structure functions $F^1_i$ these appear as divergences in the
transverse direction. 
The equation under investigation is the following:
\begin{equation}
\label{eq52}
F^{1 \rm MR}_1 + F^{1 \rm MR}_2 = F^{1 \rm R}.
\end{equation}
For the same reason as for the fermion triangle, an analytic proof of
this equation is not possible.  We investigated rotational invariance
of the left-hand side of this equation, and furthermore we checked if
it gives the same result as the covariant calculation on the right-hand
side.  A boson is at rest and decays into two fermions as indicated
in Fig.~\ref{figtheta}. The fermion
mass is taken to be the same as the boson mass. Therefore there can be
no on-shell singularities of intermediate states. Also, we dropped the
common factor~$-i$.  The contributions of the two blink diagrams are
given in the commutative diagram of Fig.~\ref{figzes}.  We made the
arbitrary choice of applying minus regularization MR1, and used 128
points in integration variable.

The error, i.e., the difference between the covariant calculation with
BPHZ regularization and the sum of minus regularized blinks, has a
maximum of 0.02\%.  This deviation results from numerical inaccuracies,
as was checked by varying the number of integration points.

We conclude that no significant deviation from a rotational invariant
amplitude is found. Moreover, we found that the sum of the light-front
time-ordered diagrams is the same as the covariant amplitude for the
one-boson exchange correction.  Again, the procedure of
$k^-$-integration and minus regularization proved to be a valid
method.

\section{Conclusions} \label{secconc}

In the Yukawa model with a scalar coupling there are five single-loop
diagrams with transverse divergences, of which two also contain
longitudinal divergences. For all other one-loop diagrams and all
multiple-loop diagrams that do not contain subdivergences, the proof of
the equivalence of covariant and light-front perturbation theory was given
by Ligterink and Bakker \cite{LB95b} upon using the $k^-$-integration
prescription.  For the two single-loop diagrams with longitudinal
divergences this integration is ill-defined. This problem was dealt
with in a previous paper \cite{SB98a}.

Of the three remaining diagrams two are thoroughly analyzed in this
paper. For the parts of these diagrams without transverse divergences
the $k^-$-integration recipe of Ligterink and Bakker applies.  For the parts with
transverse divergences a proof of equivalence is complicated by the
fact that the amplitudes depend on three independent scalar products of
the external momenta.  We applied an extended version of the method of
minus regularization invented by Ligterink and Bakker.  It is on a friendly
footing with the light-front, because it can be applied to both
longitudinal and transverse divergences.  Moreover, it has strong
similarities to BPHZ regularization, which is suitable for covariant
perturbation theory. We were able to tune the regularization in such a
way that minus regularization is analogous to BPHZ regularization. Therefore, we
expect an exact equality between the covariant and the light-front
amplitudes.  We showed that rotational invariance is maintained and we
expect that other nonmanifest symmetries on the light-front, such as
boosts in the $x$-$y$-plane, are also conserved.

The final formulas obtained did not yield to analytic integration.
Therefore we had to resort to multidimensional numerical integration.
As rotational invariance was shown previously to be violated in naive
light-cone quantization \cite{BL91}, we investigated rotational invariance,
which is one of the nonmanifest symmetries on the light-front. 
Our results demonstrate, within the errors due to the numerical methods
used, that covariant and light-front time-ordered perturbation
theory give the same physical matrix elements.

One diagram with transverse divergences has not been discussed in 
our two papers on equivalence, namely the fermion box with four external
boson lines. It is a scalar object, similar to the fermion triangle. 
The results obtained for the latter convinced us that upon 
minus regularization we shall find a covariant result. As there
are more time-orderings, and because one cannot test
for rotational invariance as easily as for the triangle diagrams,
we did not investigate this much more complicated situation. 

We trust that with our elaborate discussion of 
divergent diagrams in the Yukawa model we have illustrated the 
power of minus regularization and taken away doubts about
the covariance of light-front perturbation theory.

\section*{Acknowledgments}

The authors thank N.E.~Ligterink for discussing this work,
P.J.G.~Mulders for helpful suggestions, and A.J.~Poldervaart for
writing the first version of the {\sc fortran} code used.  This work was
supported by the Stichting voor Fundamenteel Onderzoek der Materie
(FOM), which is financially supported by the Nederlandse Organisatie
voor Wetenschappelijk onderzoek (NWO).

\appendix

\section{Internal and external variables}
\label{app1}
We get more insight into the properties of the structure functions if
we rewrite them in terms of internal and external variables.
This can be done by defining
\begin{eqnarray}
\label{a1}
x' &=& \frac{k^+}{q_1^+} = (x-1) \chi, \\
\label{a2}
x &=& \frac{k^+ + q_2^+}{q_2^+} = \frac{x' + \chi}{\chi}.
\end{eqnarray}
Or, equivalently,
\begin{eqnarray}
k^+ &=& x' q_1^+ = (x-1) q_2^+,\nonumber\\
k_1^+ &=& (x'-1) q_1^+,\nonumber\\
k_2^+ &=& x q_2^+.\nonumber
\end{eqnarray}
In the numerator of the integrals defining
light-front structure functions we encounter 
on-shell spin projections. They can be rewritten in terms of
internal variables using
\begin{eqnarray}
k^-_{1\rm on}&=& \frac{{k_1^\perp}^2 + m^2}{2 (x'-1) q_1^+},\\
k^-_{2\rm on}&=& \frac{{k_2^\perp}^2 + m^2}{2 x q_2^+}.
\end{eqnarray}
The energy denominators can also be written 
in terms of internal and external variables.
The poles are given by Eqs.~\r{pole2}, \r{pole3} and \r{pole1b}:
\begin{eqnarray}
2 q_1^+(H_1^-\!\!-\!H_2^-)=
  2 q_1^+\left(p^- \!+\! \frac{{k_1^\perp}^2 \!+\! m^2}{2k_1^+}
       - \frac{{k_2^\perp}^2 \!+\! m^2}{2k_2^+}\right)
\hspace{-.5cm}\nonumber\\
= (p^2\!\!+\!{p^\perp}^2\!) \frac{1\!+\!\chi}{\chi}
  - \frac{{k_1^\perp}^2 \!+\! m^2}{1-x'}
  - \frac{{k^\perp_2}^2\!+\! m^2}{x \chi},
\\
2 q_1^+(H_1^-\!\!-\!H^-)=
  2 q_1^+\left(q_1^-\!-\! \frac{{k^\perp}^2\!+\!\mu^2}{2k^+}
       + \frac{{k_1^\perp}^2 \!+\! m^2}{2k_1^+}\right)
\hspace{-.5cm}\nonumber\\
= q_1^2 + {q_1^\perp}^2 - \frac{{k^\perp}^2\!+\!\mu^2}{x'}
       - \frac{{k_1^\perp}^2 \!+\! m^2}{1-x'},\\
2 q_2^+(H^-\!\!-\!H_2^-)=
  2 q_2^+\left(q_2^-\!+\! \frac{{k^\perp}^2\!+\!\mu^2}{2k^+}
       - \frac{{k_2^\perp}^2 \!+\! m^2}{2k_2^+}\right)
\hspace{-.5cm}\nonumber\\
= q_2^2 + {q_2^\perp}^2 - \frac{{k^\perp}^2\!+\!\mu^2}{1-x}
       - \frac{{k_2^\perp}^2 \!+\! m^2}{x}.
\end{eqnarray}
The integration measures can be rewritten as follows:
\begin{eqnarray}
2 \pi i \int_0^{q_1^+} \frac{{\rm d}k^+ 4 q_1^+ q_2^+}{8 k_1^+ k_2^+ k^+},
&=&- \pi i \int_0^1 \frac{{\rm d}x'}{(1-x')x x'},\\
-2 \pi i \int_{-q_2^+}^0 \frac{{\rm d}k^+ 4 q_1^+ q_2^+}{8 k_1^+ k_2^+ k^+}
&=&- \pi i \int_0^1 \frac{{\rm d}x}{(1-x')x (1-x)}.
\end{eqnarray}
We conclude that it is possible to write the structure functions
in terms of the external variables
$q_1^-,\; q_2^-,\; q_1^\perp,\; q_2^\perp$ and $\chi$
and integrals over the internal variables $x$ or $x'$ and $k^\perp$.
The divergent part of the structure
functions $F^2_i$ can now be written as
\begin{eqnarray}
f^{2-}_1 =- \pi i \int_0^1 \frac{{\rm d}x'}{(1-x')x x'} \;
\frac{m}{(x'-1) q_1^+}\; \frac{q^+_1}{q^+_2} \nonumber\\
\times \left( \frac{1}{1-x'} + \frac{1}{x \chi}\right)^{-1}
\left( \frac{1}{x'} + \frac{1}{1-x'}\right)^{-1},\\
f^{2-}_2 = - \pi i \int_0^1 \frac{{\rm d}x}{(1-x')x (1-x)} \;
\frac{m}{x q_2^+} \nonumber\\
\times \left( \frac{1}{1-x'} + \frac{1}{x \chi}\right)^{-1}
\left( \frac{1}{x} + \frac{1}{1-x}\right)^{-1}.
\end{eqnarray}
Upon cancelling common factors, and using Eq.~\r{a2}, we can evaluate
the integrals and obtain
\begin{equation}
f^{2-}_1 = - f^{2-}_2 = \pi i \; \frac{\chi}{1 + \chi} \; \frac{m}{q_2^+}
= \pi i \; \frac{m}{p^+}.
\end{equation}
Therefore condition~\r{cancel} is verified.


\begin{thebibliography}{1}

\bibitem{Dir49}
P.~A.~M. Dirac,
\newblock Rev. Mod. Phys., {\bf 21}, 392 (1949).

\bibitem{KS70}
J.~B. Kogut and D.~E. Soper,
\newblock Phys. Rev. D {\bf 1}, 2901 (1970).

\bibitem{LB95b}
N.~E. Ligterink and B.~L.~G. Bakker,
\newblock Phys. Rev. D {\bf 52}, 5954 (1995).

\bibitem{LB80}
G.~P. Lepage and S.~J. Brodsky,
\newblock Phys. Rev. D {\bf 22}, 2157 (1980).

\bibitem{SB98a}
N.~C.~J. Schoonderwoerd and B.~L.~G. Bakker,
\newblock Phys. Rev. D {\bf 57}, 4965 (1998).

\bibitem{LB95a}
N.~E. Ligterink and B.~L.~G. Bakker,
\newblock Phys. Rev. D {\bf 52}, 5917 (1995).

\bibitem{Hep66}
K.~Hepp,
\newblock Commun. Math. Phys. {\bf 2}, 301 (1966).

\bibitem{HZ66}
Y.~Hahn and W.~Zimmerman,
\newblock Commun. Math. Phys. {\bf 10}, 330 (1968).

\bibitem{Zim68}
W.~Zimmerman,
\newblock Commun. Math. Phys. {\bf 11}, 1 (1968).

\bibitem{Zim69}
W.~Zimmerman,
\newblock Commun. Math. Phys. {\bf 15}, 208 (1969).

\bibitem{CK82}
W.~E.~Caswell and A.~D.~Kennedy,
\newblock Phys. Rev. D {\bf 25}, 392 (1982).

\bibitem{BL91}
M.~Burkardt and A.~Langnau.
\newblock Phys. Rev. D {\bf 44}, 3857 (1991).

\end{thebibliography}
\end{document}